\documentstyle[12pt]{article}

\large
\newcommand{\be}{\begin{eqnarray}}
\newcommand{\ee}{\end{eqnarray}}

\begin{document}
\setlength{\baselineskip}{21pt}
\pagestyle{empty}  
\vfill
\eject                                                                         
\begin{flushright}                                                             
SUNY-NTG-95-55
\end{flushright}                                                               
                                                                               
\vskip 2.0cm 
\centerline{\large {\bf Nucleons at Finite Temperature}}
\vskip 2.0 cm                                                                  
\centerline{M. Kacir and I. Zahed} 
\vskip 1cm
\vskip .5cm
\centerline{Department of Physics}
\centerline{State University of New York at Stony Brook}
\centerline {Stony Brook, New York 11794-3800}
\vskip 2cm                                                                   
                                                                               
\centerline{\bf Abstract}
The nucleon mass shift is calculated using chiral counting arguments 
and a virial expansion, without and with the $\Delta$. At all temperatures,
the mass shift and damping rate are dominated by the $\Delta$. Our results
are compared with the empirical analysis of Leutwyler and Smilga, as well
as results from heavy baryon chiral perturbation theory in the large
$N_{c}$ (number of color) limit.
We show that unitarity implies that the concepts of thermal 
shifts are process dependent.

\vfill
\noindent
\begin{flushleft}
SUNY-NTG-95-55\\
December 1995
\end{flushleft}
\eject
\pagestyle{plain}

\newpage

\section{Introduction}
In dilute many-body systems, particles undergoing 
multiple rescattering behave as quasiparticles.
 As a result, their mass and width depart from the 
vacuum values. While varying widths are a direct measurement of 
quasiparticle absorption, and may be important for
 transport properties, varying mass shifts may have fundamental
 consequences on threshold behaviours, and particle 
spectra \cite{MANY}. 

Non-perturbative model calculations at finite temperature 
and/or density support the idea that hadronic masses
 and widths change as a function of temperature and density 
\cite{QCD}. The idea that a changing rho 
mass in the hadronic medium \cite{BROWN}, may be key in understanding the 
excess of dileptons (CERES \cite{CERES}) and dimuons (HELIOS \cite{HELIOS})
below the rho mass in present relativistic heavy-ion collisions.

The definition of quasiparticle masses is, however, process dependent. 
At finite temperature a quasiparticle mass can be defined by using the 
free energy (energy mass), space-like correlation functions 
(screening mass), or time-like correlation functions (pole mass), 
to cite a few. 

The purpose of this letter is to investigate the effects of temperature on 
nucleons in a heat bath of pions. In section 2, we exploit a virial
type expression for the mass shift and the damping rate in terms of 
the forward scattering amplitude. This amplitude is obtained following
arguments based on chiral power counting arguments and on-shell Ward 
identities \cite{YAZA}.
In section 3, we give a full account of the role played by the 
$\Delta $-resonance.
A comparison with results based on empirical data is achieved
\cite{SMILGA}.
In section 4, we discuss the large $N_{c}$ limit of our results,
 then compare with recent results from heavy baryon chiral 
perturbation theory (HB$\chi $PT) \cite{BEDAQUE}.
In section 5, using a variant of the Bethe-Uhlenbeck argument,
we show that for the nucleon its mass shift is distinct from its
energy shift. In general, unitarity implies that the concepts
 of thermal shifts are process dependent.
 In section 6, we summarize our conclusions.

\section{Ward-Identity}
A nucleon immersed in a heat bath of pions undergoes rescattering 
and absorption. For temperatures $T\sim m_{\pi}$, the pion mean free 
path is about $2-3$ fm. As a result the pion thermal density is about
 $(1/3)^3$ fm$^{-3}$. The pion gas is dilute. Hence, pion-nucleon
 collisions in the heat bath may be organized in terms of the pion 
density (virial expansion). To leading order in the pion density,
 the nucleon pole mass receives a coherent shift $\Delta m_N (p)$
and damping rate $ \gamma^{T}_{N}(p)$ given by 
\begin{equation}
\Delta m_N (p) -\frac{i}{2}\gamma^{T}_{N}(p)
= -\sum_{a=1}^3 \int \frac{d^3k}{(2\pi )^3} 
\frac 1{e^{\omega_k/T}-1} \frac{{\cal T}^{aa} (p, k)}{2\omega_k}
\label{1}
\end{equation}
where ${\cal T}^{aa}$ is the forward pion-nucleon scattering amplitude,
and $\omega_k =\sqrt{k^2 + m_{\pi}^2}$. In throughout this work, we
respectively denote the pion momentum by $k$ and $q$ in the
 laboratory and center of mass frame. ${\cal T}^{aa}$ is available
 from pion-nucleon scattering data \cite{HELSINKI}, and was used by
Leutwyler and Smilga to assess the shift in the nucleon mass at zero
 momentum \cite{SMILGA}. 
Here, we will instead rely on on-shell Ward identities and chiral power 
counting to derive an expression for (\ref{1}) at tree level. This will 
help us assess the various thermal contributions to the mass and width of the
nucleon, as well as the importance of the $\Delta$.

We note that the pion-nucleon scattering
 amplitude ${\cal T}$ can be reduced 
using Weinberg's formula \cite{WEINBERG}. Taking $(k,a)$ as the 
incoming pion and $(k,b)$ the outgoing pion, the amplitude, associated
to the scattering process of Fig. 1, reads
\begin{equation}
{\cal  T} =  {\cal T}_V + {\cal T}_S +  {\cal T}_{AA}
\label{2}
\end{equation}
where
\begin{eqnarray}
{\cal T}^{ab}_V &=&
+\frac i{f_{\pi}^2} k^{\mu} 
\epsilon^{bac} \langle N(p) | {\bf V}_{\mu}^c (0) | N(p) \rangle 
\label{3} 
\\
{\cal T}^{ab}_S &=&
-\frac {m_{\pi}^2}{f_{\pi}}\delta^{ab} \langle N(p ) | 
{\bf \sigma} (0)| N(p) \rangle_{\rm conn.} = +\frac{\sigma_{\pi 
N}}{f_{\pi}^2}\delta^{ab}
\label{4}
\\
{\cal T}_{AA}^{ab} &=&
+\frac i{f_{\pi}^2} k^\mu k^\nu 
\int d^4x e^{-ik\cdot x} \langle N(p) | T^*
{\bf j}_{A\mu}^a (x) {\bf j}_{A\nu}^b (0)| N(p)
\rangle _{\rm conn.}\,\,.
\label{5}
\end{eqnarray}
Here ${\bf V}$ is the quantum vector current, ${\bf \sigma}$
 the quantum scalar density, and ${\bf j}_{A\mu}^a$ the 
quantum one-pion reduced axial-current. Their form can be found in 
Refs. \cite{YAZA,JIM} for QCD and sigma models. In the forward 
amplitude (\ref{2}), the isovector charge of the nucleon (\ref{3})
 drops, and one is left with the sigma-term contribution 
(\ref{4}) and the one-pion reduced axial-axial 
correlator (\ref{5}). Thus
\begin{equation}
\sum_{a=1}^3{\cal T}^{aa} =
 3\frac{\sigma_{\pi N}}{f_{\pi}^2} +
\frac i{f_{\pi}^2} k^\mu k^\nu 
\int \! d^4x\, e^{-ik\cdot x} \sum_{a=1}^3\langle N(p) | T^*
{\bf j}_{A\mu}^a (x) {\bf j}_{A\nu}^a (0)| N(p)\rangle _{\rm conn.}
\label{7}
\end{equation}
Inserting (\ref{7}) into (\ref{1}) we obtain for the mass shift and
damping rate
\begin{eqnarray}
\Delta m_N (p) -\frac{i}{2}\gamma_{N}^{T}(p)
= &-&3n_{\pi} (T)\,\,\frac{\sigma_{\pi N}}{f_{\pi}^2} -
\sum_{a=1}^3 \int \frac{d^3k}{(2\pi )^3} \frac 1{2\omega_k}
\frac 1{e^{\omega_k/T}-1} \nonumber\\&\times&
\frac i{f_{\pi}^2} k^\mu k^\nu 
\int \! d^4x\, e^{-ik\cdot x} \langle N(p) | T^*
{\bf j}_{A\mu}^a (x) {\bf j}_{A\nu}^a (0)| N(p)\rangle _{\rm conn.}
\label{8}
\end{eqnarray}
where 
\begin{equation}
3 n_{\pi } (T) = \sum_{a=1}^3 \int \frac{d^3k}{(2\pi )^3} 
\frac{1}{2\omega_k}
\frac 1{e^{\omega_k/T}-1} =
\sum_{a=1}^3 \int \frac{d^4k}{(2\pi )^4} \, 2\pi \,
 \delta (k^2-m_{\pi}^2)\,
\frac 1{e^{|k^0|/T}-1} 
\label{9}
\end{equation}
counts the number of pions per unit energy and per unit volume
in the heat bath. For massless pions,
 this is just $T^2/4$. To assess the role of the one-pion reduced
 axial-axial correlator, we will use  power counting in $1/f_{\pi}$
as discussed in Refs. \cite{YAZA,JIM} with and without the $\Delta$.

In the absence of $\Delta$, the one-pion reduced axial-current at 
tree level reads \cite{JIM}
\begin{equation}
{\bf j}_{A\mu}^a 
= g_A{\overline{\bf \Psi}}\gamma_{\mu} \gamma_5
\frac{\tau^a}{2} {\bf \Psi} 
+\frac {\sigma_{\pi N}}{m_{\pi}^2} \partial_{\mu}
\left( \overline{\bf \Psi} i\gamma_5 \,
 \tau^{a} {\bf \Psi} \right)
+{\cal O} \left(\frac 1{f_{\pi}^2}\right)
\label{10}
\end{equation}
where $g_A$=1.357 is the nucleon axial-coupling and ${\bf \Psi}$ is 
the nucleon-field. We take $m_{\pi}$=139 MeV and $m_{N}$=939 MeV for
the pion and nucleon mass, the empirical value $\sigma _{\pi N}=
45\pm 8$ MeV and the pion decay constant $f_{\pi}=$94 MeV.
 Our expansion is on-shell, so all the 
parameters in our expressions are set at their physical (on mass-shell 
renormalization). The corrections follow from pion and nucleon loops 
\cite{JIM}. Inserting (\ref{10}) into the scattering amplitude 
(\ref{5}) yields
\begin{eqnarray}
{\cal T}_N (\nu )&=&\sum_{a=1}^3{\cal T}^{aa}_{AA}
\nonumber \\
&=& - \frac {3}{4f_{\pi}^2 }\bigg(
4g_A (2\sigma_{\pi N} -g_Am_N) 
-\frac{(\sigma_{\pi N} - g_A m_N )^2}{2m_N}
\left( \frac{4\nu }{\nu -\nu_N} +
\frac{4\nu}{\nu +\nu_N} \right) \bigg)
\nonumber \\
& + &{\cal O} \left( \frac 1{f_{\pi}^4} \right)
\label{11x}
\end{eqnarray}
where $\nu=p\cdot k/m_{N}=(E_{q}\omega _{q}+q^{2})/m_{N}$
 and $\nu_N =m_{\pi}^2/2m_N$.
If we note that to this order the 
pion-nucleon S-wave scattering length is given by
\begin{equation}
4\pi\left(1+\frac{m_\pi}{m_N}\right) a^+ = 
\frac{\sigma_{\pi N}}{f_{\pi}^2}
+ \frac 13 {\cal T}_N (m_{\pi} )
\label{12}
\end{equation}
then the nucleon mass shift and damping rate (\ref{8}) read to order
${\cal O}(1/f_{\pi}^2)$, in terms of real and imaginary parts of the 
scattering amplitude
\begin{eqnarray}
\Delta m_N (0) =&-& 3 n_{\pi} (T) \left(1+ \frac {m_{\pi}}{m_N}\right) 
4\pi a^+\nonumber\\
&-&
\int \! \frac{d^3k}{(2\pi^3)} \,\frac{ 1}{2\omega_k}
 \frac{ 1}{e^{\omega_k/T}-1}
{\rm Re}
\left( {\cal T}_N (\omega_k ) - {\cal T}_N (m_{\pi}) \right) 
+{\cal O} \left(\frac{1}{f_{\pi}^4}\right)
\label{13}
\end{eqnarray}
and
\begin{equation}
\gamma_{N}^{T}(0)=
2\int \! \frac{d^3k}{(2\pi^3)}\, \frac{1}{2\omega_k}
\frac{1}{e^{\omega_k/T}-1}
{\rm Im}
\left( {\cal T}_N (\omega_k ) - {\cal T}_N (m_{\pi}) \right)
 +{\cal O} \left(\frac{1}{f_{\pi}^4}\right)
\label{13a}
\end{equation}
This is an exact result of chiral symmetry. At tree level, and in 
the absence of the isobar, there is no absorption and the damping
 rate vanishes.  
Each contribution from both the scattering length term and the nucleon
is displayed in Fig. 3. For temperatures  $T< m_{\pi}$, they are
opposite in sign and of the order of 1 MeV resulting in a 
nucleon mass shift which is negative and less than 1 MeV over 
this range of temperature. 
 
\section{$\Delta $ resonance }
 The role of the $\Delta$ with a finite width cannot be assessed 
uniquely in the formalism developed in Refs. \cite{YAZA,JIM}. 
Given its importance in the axial-axial correlator, we will try
 to approximately assess its contribution in this section. 
We will assume that the $\Delta$ has a zero width 
(exact argument) and then assign by  hand a 
width in the final result (approximate argument). 
With this in mind, the pion-nucleon-delta
coupling will be taken to be pseudovector with strength,
\begin{equation}
\frac {g_{\pi N\Delta}}{2m_{\Delta}}\,\, 
\left(\partial_{\mu}\pi^{a}\right)\,\,
\left(
\overline{\bf \Psi}\, {\bf Q}^{\mu, a}\, +\,{\it h.c.}
\right)
\label{vertex}
\end{equation}
where $\pi^a$ is the quantum PCAC pion field \cite{YAZA,JIM}, and
${\bf Q}^{\mu, a}$ is a Rarita-Schwinger field \cite{LURIE} with
 vector index $\mu$ and isospin index $a$. The (omitted)
Dirac and isodoublet indices are contracted over with the
 nucleon field. The transition matrix element between the nucleon
 with momentum $p$ and isospin $b$, and the isobar with momentum
 $p'$ and isospin $b'$, induced by the one-pion 
reduced axial-vector current takes the general form
\begin{eqnarray}
\langle N(p, b) |\,{\bf j}_{A\mu}^a (0)\,| 
\Delta (p', b') \rangle&=& 
\overline{u} (p, b) \bigg( F (k^2) g_{\mu\nu} + G(k^2) 
\gamma_{\mu} k_{\nu}
\nonumber\\& &\ 
+ H(k^2) k_{\mu} k_{\nu} + i I (k^2) 
\sigma_{\mu}^{\lambda} k_{\lambda} 
k_{\nu}\bigg) \,\,{\bf U}^{\nu, a} (p', b')
\label{16}
\end{eqnarray}
where $k^2=(p-p')^2$ and ${\bf U}$ is a Rarita-Schwinger spinor. 
The general structure (\ref{16}) follows from Lorentz invariance,
 P- and T-invariance \cite{SOLDATE}. At tree level, the form 
factors obey a Goldberger-Treiman relation
\begin{equation}
\frac{ g_{\pi N\Delta}}{2m_{\Delta}} 
= \frac{ F + G (m_{\Delta} -m_N ) }{f_{\pi}}
\label{17}
\end{equation}
where $F = F(0)$ and $G = G(0)$. The contribution of the $\Delta$ 
to the one-pion 
reduced axial-axial correlator in the nucleon (\ref{8}) can be
 evaluated using (\ref{16}) and the following $\Delta$ propagator
\begin{eqnarray}
{\bf \Delta}_{\mu\nu}^{ab}(p) = &&(\delta^{ab} - 
\frac{1}{3} \tau^a\tau^b )
\frac 1{\rlap/p -m_{\Delta} }\nonumber\\
&\times&
\left( -g_{\mu\nu} +\frac{1}{3} \gamma_{\mu}\gamma_{\nu} 
- \frac{2}{3m_{\Delta}^2} p_{\mu} p_{\nu} 
+\frac{1}{3m_{\Delta}} (\gamma_{\mu} 
p_{\nu}-p_{\mu} \gamma_{\nu} )\right)
\label{18}
\end{eqnarray}
In Fig. 2, we exhibit the scattering process considered in the heat
bath.
The resulting amplitude is
\begin{eqnarray}
{\cal T}_{\Delta} (\nu ) =&-&\frac{4}{3f_{\pi}^2}
\,\Lambda (\nu)
\,\bigg( -G^{2}(\nu+m_{\Delta}-m_{N})+ \frac{g_{\Delta}^{2}
}{2m_N} 
\frac{ \nu + m_{\Delta} + m_N }{\nu-\nu_{\Delta}}\bigg)
\nonumber \\
&+&(\nu \rightarrow -\nu)\, +\,
{\cal O}\left( \frac{1}{f_{\pi}^4}\right)
\label{ampdelta}
\end{eqnarray}
where $\nu_{\Delta}=(m_{\Delta}^2-m_N^2-m_{\pi}^2 )/2m_N$, 
  $g_{\Delta}^{2}=(F^2- G^2 (m_{\Delta} -m_N)^2)$ and
\\
\begin{equation}
\Lambda (\nu)= m_{\pi}^{2}
-\frac{(m_{N}\nu+m_{\pi}^{2})^{2}}{m_{\Delta}^{2}}
\end{equation}
\\
This is to be compared with  
the nucleon contribution (\ref{11x}) discussed above.

In terms of (\ref{11x}) and (\ref{ampdelta}) the contribution to 
order ${\cal O} (1/f_{\pi}^4)$ to the isospin averaged
 pion-nucleon scattering amplitude (\ref{7}) reads
\begin{equation}
\sum_{a=1}^3 {\cal T}^{aa} (\nu) = 
12\pi a^{+} (1+ \frac {m_{\pi}}{m_N} )
+ \left( {\cal T}_N (\nu) + {\cal T}_{\Delta} (\nu ) 
-{\cal T}_{N} (m_{\pi})-{\cal T}_{\Delta} (m_{\pi} ) \right) 
\label{totalamp}
\end{equation}
where the S-wave scattering length now receives a contribution from 
the isobar through
\begin{equation}
4\pi a^{+} (1+\frac {m_{\pi}}{m_N} ) 
= \frac {\sigma_{\pi N}}{f_{\pi}^2}
+\frac{1}{3} {\cal T}_{N} (m_{\pi}) 
+\frac{1}{3} {\cal T}_{\Delta} (m_{\pi})
\label{22}
\end{equation}
In terms of the real and imaginary parts of the scattering amplitude
in (\ref{totalamp}), the nucleon mass shift is given by
($p=0$ is understood)
\begin{eqnarray}
\Delta m_{N} =&-&3 n_{\pi} (T) 
\left(1+ \frac{m_{\pi}}{m_{N}}\right) 
4\pi a^+ \nonumber\\
&-&
\int\! \frac{d^3k}{(2\pi^3)}\, \frac{1}{2\omega_k} 
\frac{1}{e^{\omega_k/T}-1}
{\rm Re}
\bigg( {\cal T}_{N} (\omega_k) 
+ {\cal T}_{\Delta} (\omega_k ) 
-{\cal T}_{N} (m_{\pi})
-{\cal T}_{\Delta} (m_{\pi} ) \bigg) 
\nonumber \\
&+&{\cal O} \left( \frac{1}{f_{\pi}^4} \right)
\label{23}
\end{eqnarray}
and the damping rate by
\begin{eqnarray}
\gamma_N ^{T} =&+&
2\,\int \! \frac{d^3k}{(2\pi^3)}\, 
\frac{1}{2\omega_k} \frac{1}{e^{\omega_k/T}-1}
{\rm Im}
\bigg( {\cal T}_{N} (\omega_k) 
+ {\cal T}_{\Delta} (\omega_k ) -{\cal T}_{N} 
(m_{\pi})-{\cal T}_{\Delta} (m_{\pi} ) \bigg) 
\nonumber \\
&+&{\cal O} \left( \frac{1}{f_{\pi}^4} \right)
\label{23a}
\end{eqnarray}
This result relies on the choice of the interaction (\ref{vertex}),
 and as such is not unique.
 Since the $\Delta$ sits in the continuum, (\ref{23}) diverges. 
In nature, $\Delta$ has a width $\Gamma_{\Delta}$ and
 (\ref{23}) is finite.
To account for this width, we parameterize the resonant part of 
the scattering amplitude (\ref{18}) in the P33 channel by a
Breit-Wigner form.

To this end, we decompose the amplitude (\ref{ampdelta}) as follows
\cite{HOHLER}
\begin{equation}
{\cal T}_{\Delta}(\nu)=3(A^{+}_{\Delta ,p}+\nu B^{+}_{\Delta ,p}
+A^{+}_{\Delta ,np}+\nu B^{+}_{\Delta ,np})
\label{new1}
\end{equation}
where the pole parts
\begin{equation}
A^{+}_{\Delta ,p}(\nu) 
= A^{+}_{\Delta ,dp}(\nu) +A^{+}_{\Delta ,cp}(\nu) 
=  \frac{g_{\Delta}^{2}}{9m_{N}}
\, \alpha_1
\bigg(\frac{1}{\nu_{\Delta}-\nu} + \frac{1}{\nu_{\Delta}+\nu}\bigg)
\label{ap}
\end{equation}
\begin{equation}
B^{+}_{\Delta ,p}(\nu) =B^{+}_{\Delta ,dp}(\nu)
+B^{+}_{\Delta ,cp}(\nu) =
\frac{g_{\Delta}^{2}}{9m_{N}}
\, \beta_1
\bigg(\frac{1}{\nu_{\Delta}-\nu} 
- \frac{1}{\nu_{\Delta}+\nu}\bigg)
\label{bp}
\end{equation}
and the non-pole parts
\begin{eqnarray}
A^{+}_{\Delta ,np}(\nu) =
&-&\frac{4g_{\Delta}^{2}}{9m_{\Delta}}
\bigg((E_{\Delta}+m_{N})(2m_{\Delta}-m_{N})
+m_{\pi}^{2}(2+m_{N}/m_{\Delta})
\bigg)
\nonumber \\
&+&\frac{4G^{2}}{9f_{\pi}^{2}}\,
2(m_{\Delta}-m_{N})
\bigg(m_{\pi}^{2}+\frac{m_{N}^{2}\nu
^{2}+m_{\pi}^{4}}{m_{\Delta}^{2}}\bigg)
\label{anp}
\end{eqnarray}
\begin{equation}
B^{+}_{\Delta ,np}(\nu) =
-\frac{4g_{\Delta}^{2}}{9m_{\Delta}}\frac{m_{N}\nu}{m_{\Delta}}
+\frac{16G^{2}}{9f_{\pi}^{2}}\,\frac{m_{N}\nu
m_{\pi}^{2}}{m_{\Delta}^{2}}
\label{bnp}
\end{equation}
with
\begin{eqnarray}
g^{2}_{\Delta}&=&\bigg(F^{2}-G^{2}(m_{\Delta}
-m_{N})^{2}\bigg)/f_{\pi}^{2}
\nonumber \\
\alpha_1 & = &3(m_N +m_{\Delta}) (E_{\Delta}^2-m_N^2) + 
(m_{\Delta}-m_N)(E_{\Delta}+m_N)^{2}
\nonumber\\
\beta_1 &=& 3 (E_{\Delta}^2-m_N^2)-(E_{\Delta}+m_N)^2
\nonumber\\
E_{\Delta} \pm m_N &=&\frac 1{2m_{\Delta}} 
\left( (m_{\Delta} \pm m_N)^2 -m_{\pi}^2\right)
\label{new2}
\end{eqnarray}
The procedure for introducing the width of the $\Delta$ follows the
approach in \cite{HOHLER2} and is discussed
in details in the appendix.

Using the empirical values $\sigma_{\pi N} = 45\pm 8$ MeV 
\cite{SAINIO}, $a^+ = -(8\pm 4)\ 10^{-3}/m_{\pi}$ \cite{HELSINKI},
 and the SU(3) relation
 $g_{\pi N\Delta} = 3g_{\pi NN}/\sqrt{2}$,
 the Goldberger-Treiman relation (\ref{17}) and the expression
(\ref{22}) of the scattering length $a^{+}$ in terms of threshold
amplitudes allow for a determination of the isobar
 axial-form factors $F$ and $G$. 
We have examined the resulting amplitude opposite to the
experimental data \cite{HOHLER} and noticed a very poor fit at low
energy.
Instead, we rely on the empirical values at threshold of
 $A^{+}(q=0)=227.3$ GeV$^{-1}$ and
$B^{+}(q=0)=-1639$ GeV$^{-1}$.
The results are 
$F =1.382$ and $G= 4.235\ 10^{-4} $ MeV$^{-1}$. 
The first is to be compared with $F=3g_A/2\sqrt{2}\sim 
1.31$ from large $N_c$ arguments \cite{SOLDATE}. 
In Figs. 4 and 5, we display both real and imaginary parts of  
$A^{+}(\nu)$ and $B^{+}(\nu)$ opposite the empirical data from
\cite{HOHLER}. 

The mass shift of the $\Delta$ follows from (\ref{23}) after
 inserting (\ref{totalamp}) and re-expressing the phase space 
integral from the nucleon rest frame, to the center of mass frame.
 The various contributions to the nucleon 
mass shift, stemming from the scattering length, the nucleon Born 
term, the $\Delta$ Born term (with width), and the combined sum,
  are shown in Fig. 3.
In Fig. 6, the total mass shift $\Delta m_{N}$ (full line)
 is compared to the one (dotted line)
obtained from the empirical scattering 
amplitude \cite{HOHLER}. 
 The rise in the nucleon mass at temperatures of 
the order of the pion mass is conditioned by the resonance. 
The width of the $\Delta$ yields an imaginary part to the mass 
shift. Nucleons in the heat bath undergo strong absorption through 
$\pi N\rightarrow \Delta$.  
In Fig. 7, we display the damping rate $\gamma _{N}^{T}$ (full line)
opposite to the empirical one (dotted line).
We point out that the nucleon mass shift was obtained in the same
manner by Leutwyler and Smilga \cite{SMILGA}. For the damping rate,
use of the total $\pi\ p$ cross section was made in \cite{SMILGA}.
 Overall, we can observe that our chiral counting argument for the
scattering amplitude allows us to reproduce quantitatively 
both the nucleon mass
shift and damping rate up to temperatures of the order of m$_{\pi}$.
As already pointed out in \cite{SMILGA} for higher temperatures, 
we would expect other contributions like the N$^{*}$ to play a role.
In this regime, however, the pion gas is no longer dilute and the 
use of the first two terms in the virial expansion is no longer 
justified. 

\section{Large $N_{c}$ limit}

In this section, we will examine the forward scattering amplitude
in the large $N_{c}$ limit. 
As we will see below, the contribution of the width being subleading
in $N_{c}$, we will ignore it in the discussion for the mass shift
$\Delta m_{N}$.
With $m_{N},\ m_{\Delta},\ \sigma _{\pi N},\ g_{A},\ f_{\pi }^{2},
\ F $ and $G$
all of order ${\cal O}(N_{c})$ and denoting $\mu =m_{\Delta}-m_{N}
\sim {\cal O}(1/N_{c})$, we can write for the forward scattering
amplitude for the nucleon
\begin{equation}
{\cal T}_{N}(\nu)\sim
\frac{3g_{A}^{2}m_{N}}{f_{\pi}^{2}}\,
\frac{\nu_{N}^{2}}{\nu ^{2}-\nu _{N}^{2}}
\sim {\cal O}(N_{c}^{0})
\end{equation}
and for the $\Delta$
\begin{equation}
{\cal T}_{\Delta}(\nu)\sim
\frac{4}{3f_{\pi}^{2}}\,
\nu ^{2}
\bigg(-2G^{2}\mu+\frac{F^{2}}{m_{N}}
+\frac{F^{2}}{m_{N}}\, \frac{2m_{N}\mu}{\nu ^{2}-\mu ^{2}}\bigg)
\sim 
{\cal O}(N_{c}^{0})
\end{equation}
The contribution from the sigma term is
\begin{equation}
{\cal T}_{\sigma}(\nu)\sim
\frac{3\sigma _{\pi N}}{f_{\pi}^{2}}
\sim 
{\cal O}(N_{c}^{0})
\end{equation}
Setting $m_{\pi}=0$, we express the mass shift
$\Delta m_{N} =\Delta m_{N,np} +\Delta m_{N,p} $ where the subscripts
$np$ and $p$ respectively indicate non-pole and pole contributions.
The non-pole part reads
\footnote{Note that the term $\mu /T$ in $\Delta m_{N,p}$, is subleading in
$1/N_c$ and should be effectively dropped. We have kept it to make the 
comparison with \cite{BEDAQUE} immediate.}
\begin{equation}
\Delta m_{N,np}=
-\frac{T^{2}}{4\pi ^{2}f_{\pi}^{2}}
\int _{0}^{\infty}\!\frac{x\,dx}{e^{x}-1}
\, \bigg(3\sigma _{\pi N}-\frac{4}{3}x^{2}T^{2}
\left(-2G^{2}\mu+\frac{F^{2}}{ m_{N}}\right)\bigg)
\end{equation}
The pole part reads
\begin{equation}
\Delta m_{N,p}=
\frac{2F^{2}}{3}\,\frac{\mu T^{2}}{\pi^{2} f_{\pi}^{2}}
\,\int _{0}^{\infty}\!dx\ \frac{x^{3}}{x^{2}-\mu ^{2}/T^{2}}
\frac{1}{e^{x}-1}
\end{equation}
Both pole and non-pole contributions are of order 
${\cal O}(N_{c}^{0})$.
The mass shift obtained can be compared to the result quoted in
\cite{BEDAQUE}, where heavy baryon chiral perturbation theory
(HB$\chi $PT) arguments were used.
The non-pole term is absent in \cite{BEDAQUE} and the result quoted 
for the pole term has $2F^{2}/3$ substituted by $\chi _{N}g^{2}$
where $\chi _{N}=15$ and $g=3(F+D)/5$.
This mismatch in the factor $2F^{2}/3$ and the absence of a non-pole
term may explain a mass shift with essentially the same
trend as ours, however from \cite{BEDAQUE}, $\Delta m_{N}$ is larger
than zero for $T>80 $ MeV.
In Fig. 8, we display the pole and non-pole contributions to the 
mass shift. 
 The short-dashed line represents the sum of the pole and non-pole
 contribution and the long-dashed line the mass shift from chiral
 power counting of Fig. 6.

In order to examine the damping rate in the large $N_{c}$ limit,
we first write the P33 wave for the direct term in the forward
scattering amplitude (see appendix)
\begin{equation}
f_{33}^{P}\sim\frac{F^{2}}{12\pi f_{\pi}^{2}}
\, \frac{q^{2}}{\nu_{\Delta} -\nu}
\end{equation}
Using a Breit-Wigner form to describe the width of the $\Delta$ and 
keeping in mind that this is valid at resonance, the imaginary part of
the P33 wave reads
\begin{equation}
{\rm Im }f_{33}^{P}=\frac{\Gamma _{\Delta}}{2q}
\, \frac{\Gamma _{\Delta}/2}
{(\nu _{\Delta} -\nu)^{2}+(\Gamma _{\Delta }/2)^{2}}
\label{ima33}
\end{equation}
where, in the large $N_{c}$ limit, the width reads
\begin{equation} 
\Gamma _{\Delta}=\frac{F^{2}}{6\pi f_{\pi}^{2}}\, \mu ^{3}
\sim {\cal O}(1/N_{c}^{2})
\label{ncwidth}
\end{equation}
Due to the order of $\Gamma _{\Delta}$ in (\ref{ncwidth}), we can
further write the P33 wave
\begin{equation}
{\rm Im }f_{33}^{P}=\frac{\Gamma _{\Delta}}{2q}
\, \pi \delta(q-\mu)
\label{ncf33}
\end{equation}
 It is remarkable that the leading term in $N_{c}$
for ${\rm Im }f_{33}^{P}$ is exactly the result obtained
from the pole approximation approach to the width of the $\Delta$
as a narrow resonance \cite{pole}.
With this in hand, it is not difficult to obtain the leading term 
in $N_{c}$ for the imaginary part of the forward scattering amplitude.
The result is
\begin{equation}
{\rm Im}{\cal T}_{\Delta}(k)=\frac{4\pi}{3}\, 
\frac{F^{2}}{f_{\pi}^{2}}\, \mu ^{2}\delta(k-\mu)
\end{equation}
The damping rate follows from (\ref{23a}) and is of order 
${\cal O} (1/N_{c}^{2})$
\begin{equation}
\gamma _{N}^{T}=\frac{2F^{2}}{3}\, \frac{\mu ^{3}}{\pi f_{\pi}^{2}}
\frac{1}{e^{\mu /T}-1}
\label{dpnc}
\end{equation}
Again, if we attempt a comparison with \cite{BEDAQUE}, we need
to substitute $2F^{2}/3$ by $\chi _{N} g^{2}/12$.
In Fig. 9, we display the damping rate (\ref{dpnc}) (full line) 
opposite the result obtained from chiral power counting (long-dashed
line).

The present large $N_c$ results for the nucleon mass shift and damping rate,
can be effectively tested in the Skyrme model. This point will be discussed 
elsewhere \cite{KACIRZAHED}. 

\section{Process Dependence }
The above definition of the mass shift relies on the pole mass
definition. Is it definition (process) independent ?
 In this section we will show that in general unitarity forces
 the concept of thermal shifts to be process dependent,
 weakly when "perturbative" cuts are involved and strongly
when "resonant" cuts are involved. Our argument is generic, 
although we will use the nucleon for illustration.

Let $\Delta E_N$, be the shift in the energy of a single nucleon
 immersed in a dilute heat bath of pions.
 To first order in the pion density, the shift is 
\be
\Delta E_{N}=\int_{0}^{+\infty}\frac{dk}{\pi}\sum _{I,l} \, (2l+1)
\delta _{I,l}^{\prime}(k) \frac{\omega_k}{e^{\omega_k/T}-1}
\label{26}
\ee
where $\delta_{l,I}$ is the phase-shift of a pion partial 
wave carrying angular momentum $l$ and isospin $I$,
 expressed in the rest frame of the nucleon. 
(\ref{26}) is just the thermal zero-point energy if we recall that 
in a spherical box of size $R$, the boundary condition on the lth 
partial wave is
\begin{equation}
kR -l\frac{\pi}2 +\delta_{I,l} (k) = n\pi
\label{new5}
\end{equation}
where $n$ is integer. Hence, the change in the number of states 
per unit $k$ is $dn/dk = \delta_{I,l}' (k)/\pi$. 
Hence (\ref{26}). In terms of the partial phase shifts
 $\delta_{I,l}$, the scattering amplitude ${\cal T}_I (k)$ for 
fixed isospin $I$ (our ${\cal T}^{aa}$ above), is given by the
 conventional spherical expansion
\begin{equation}
{\cal T}_I (k) =i\frac {2\pi}k \sum_{l} (2l+1) 
\left(e^{2i\delta_{I,l} (k)} -1\right) \,\,P_l (\hat k )
\label{new6}
\end{equation}
Using (\ref{new6}), we can rewrite (\ref{26}) in terms of the
 scattering amplitude, a procedure known from the work of Bethe and 
Uhlenbeck \cite{LANDAU}. The result is
\begin{eqnarray}
\Delta E_N =&+&\sum_I\int\! \frac{d^3k}{(2\pi )^3}\, 
\frac {{\rm Re}{\cal T}(k)}{2\omega_k}
\left( \frac 1{e^{\omega_k/T} -1} - \frac {\omega_k}T 
\frac {e^{\omega_k/T}}{(e^{\omega_k/T}-1)^2}\right) \nonumber\\
&&-\frac{i}{8\pi}\sum_I\int \! \frac{d^3k}{(2\pi)^3}\,
\left({\cal T}_I {\cal T}_I^{* '} -{\cal T}_I^*{\cal T}_I' \right) (k)
\frac{\omega_k}{e^{\omega_k/T}-1}
\label{29}
\end{eqnarray}
A comparison between (\ref{1}) and (\ref{29}) shows that the real 
part of the mass shift is in general different from the thermal energy
shift, for complex ${\cal T}$. 
This is always the case if the amplitude satisfies unitarity. At low 
temperatures, however,
the pion unitarity cuts are suppressed by powers of $(T/f_{\pi})$,
 and one may argue that the tree calculations may be sufficient. 
Hence that a thermal energy shift becomes a pole mass shift 
(their difference being an entropy-like term), both of which real.
 This argument, however, is incorrect given the nearness
 of the $\Delta$ resonance to the pion-nucleon threshold,
and we conclude that the concept of mass shift for the nucleon is process 
dependent.

\section{Conclusion}
We have analyzed the effects of temperature on the nucleon mass, 
using the pole-mass definition.
 To leading order in the density, the mass shift and damping rate
were determined exactly to order $1/f_{\pi}^2$ 
using chiral counting arguments with and without the isobar.
 For temperature $T$ below the pion mass $m_{\pi}$ where the
 validity of the virial expansion holds, the role of the
 $\Delta $-resonance has proved to be the crucial contribution 
in the mass shift and damping rate of the nucleon.
 Other effects, like the S-wave scattering length and the one-pion
coupling to the nucleon, yield small contributions (of the order
of a few percent) compared to this resonance.
 This result, is in agreement with a virial
 calculation using the pion-nucleon scattering data \cite{SMILGA}, 
and sum rule calculations \cite{ADAMI}. 
A large $N_{c}$ investigation of our results has confirmed the results
obtained for the mass shift and damping rate in a heavy baryon chiral
perturbation theory (HB$\chi $PT) \cite{BEDAQUE}.
We have shown that unitarity implies 
that the concept of a mass shift is process dependent.

\section {Acknowledgements}

This work was supported in part by a US DOE grant DE-FG-88ER40388.

\newpage
\appendix{\large{{\bf Appendix}}}

In this appendix, we elaborate on how we introduce 
the width of the $\Delta$
resonance. We adopt the notations of \cite{HOHLER} throughout 
this appendix.
We start with the partial wave decomposition of the direct
term in $1/(\nu _{\Delta} -\nu)$ in the pole part of the 
invariant amplitude in (\ref{ampdelta})
\begin{equation}
A_{l}^{\pm}(s)=\int _{-1}^{1}\!dz\, P_{l}(z)
A_{\Delta ,dp}^{\pm}(s,t)
\end{equation}
where $z=\cos\theta$, $\theta$ being the scattering angle in 
the center of mass frame. $s$ and $t$ are the usual Mandelstam 
variables and $P_{l}(z)$ the Legendre polynomial of order $l$.
We can also write a similar expression for $B_{l}^{\pm}(s)$. 
We note that the $t$ dependence in the amplitude is kept in order
 to correctly project out the waves. This requires the respective
 change $\alpha _{1} \rightarrow \alpha _{1}+\alpha _{2}t$ and
$\beta _{1} \rightarrow \beta _{1} +\beta _{2}t$ in (\ref{ap}) 
and (\ref{bp}) where
$\alpha _{2}=3q_{\Delta}^{2}-(E_{\Delta}+m_{N})^{2}$ and $\beta _{2}=
3/2$. 
We obtain
\begin{equation}
A_{0}^{\pm}(s)\,=\,\left(\begin{array}{c} 2 \\ -1 
\end{array}\right)\,
\frac{g_{\Delta}^{2}}{18m_{N}}\,2\,
(\alpha _{1}-2\alpha _{2}q^{2})\,
\frac{1}{\nu_{\Delta}-\nu}
\nonumber 
\end{equation}
and
\begin{equation}
A_{1}^{\pm}(s)\,=\,\left(\begin{array}{c} 2 \\ -1 
\end{array}\right)\,
\frac{g_{\Delta}^{2}}{18m_{N}}\,\frac{2}{3}\alpha _{2}q^{2}\,
\frac{1}{\nu_{\Delta}-\nu}
\nonumber 
\end{equation}
The respective expressions for $B_{0}^{\pm}(s)$ 
and $B_{1}^{\pm}(s)$ are obtained from $A_{0}^{\pm}(s)$ and 
$A_{1}^{\pm}(s)$ with $\alpha _{1,2} \rightarrow \beta _{1,2}$.
All other higher waves vanish.
Furthermore, we now write the partial waves of the spin no-flip and
spin flip of the full amplitude as follows (the $s$ dependence is
understood) 
\begin{eqnarray}
\frac{32\pi}{3}\sqrt{s}\,f_{31}^{S}=&+&
(E+m_{N})\bigg(A_{0}^{+}+(\sqrt{s}-m_{N})B_{0}^{+}\bigg)
\nonumber \\
&+&(E-m_{N})\bigg(-A_{1}^{+}+(\sqrt{s}+m_{N})B_{1}^{+}\bigg)
\nonumber 
\end{eqnarray}
\begin{eqnarray}
\frac{32\pi}{3}\sqrt{s}\,f_{31}^{P}=&&
(E+m_{N})\bigg(A_{1}^{+}+(\sqrt{s}-m_{N})B_{1}^{+}\bigg)
\nonumber \\ 
&+&(E-m_{N})\bigg(-A_{0}^{+}+(\sqrt{s}+m_{N})B_{0}^{+}\bigg)
\nonumber 
\end{eqnarray}
\begin{equation}
\frac{32\pi}{3}\sqrt{s}\,f_{33}^{P}=
(E+m_{N})\bigg(A_{1}^{+}+(\sqrt{s}-m_{N})B_{1}^{+}\bigg)
\nonumber 
\end{equation}
\begin{equation}
\frac{32\pi}{3}\sqrt{s}\,f_{33}^{D}=
(E-m_{N})\bigg(-A_{1}^{+}+(\sqrt{s}+m_{N})B_{1}^{+}\bigg)
\label{spinflip}
\end{equation}
The width of the isobar is constructed through a Breit-Wigner form
for the P33 channel, namely
\begin{equation}
f^{P}_{33} (q ) 
\rightarrow
\frac{f^{P}_{33} (q )} 
{1\,-\,iq\cdot f^{P}_{33} (q )}
\label{new3}
\end{equation} 
With this in hand, we reconstruct the pole part of the amplitude
$A^{+}_{\Delta,dp}(\nu)$ ($B^{+}_{\Delta,dp}(\nu)$)
by inverting the equations for each channel $\alpha $ in $f_{\alpha}$
 (\ref{spinflip}).
In turn, we can combine these terms with the crossed terms of the pole
part $A^{+}_{\Delta,cp}(\nu)$ ($B^{+}_{\Delta,cp}(\nu)$), the non-
pole part $A^{+}_{\Delta,np}(\nu)$ ($B^{+}_{\Delta,np}(\nu)$) and the
nucleon contribution $A^{+}_{N}(\nu)$ ($B^{+}_{N}(\nu)$)
In Figs. 4a and 4b, we respectively display in full line
the real parts ${\rm Re}\, A^{+}$ and ${\rm Re}\, B^{+}$
 and in Figs. 5a and 5b the imaginary parts
${\rm Im}\, A^{+}$ and ${\rm Im}\, B^{+}$ 
opposite to the experimental data points \cite{HOHLER}. Our
procedure is artificial in the sense that it mimics the data in the
low energy regime. For instance, the width of the $\Delta $ turns
 out to be $\Gamma _{\Delta}\sim 90$ MeV with the mass 
$m_{\Delta}=1220$ MeV. This is in contrast with the known values 
$\Gamma _{\Delta}\sim 114$ MeV and $m_{\Delta}=1232$ MeV.

\newpage
{\large{\bf Figure captions }}

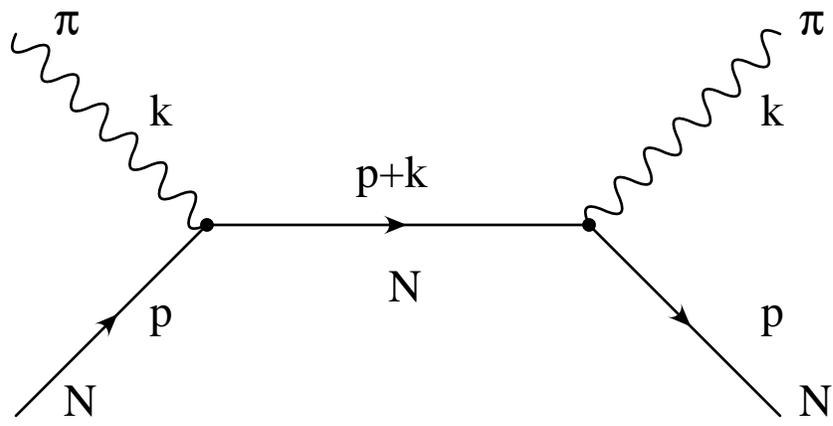
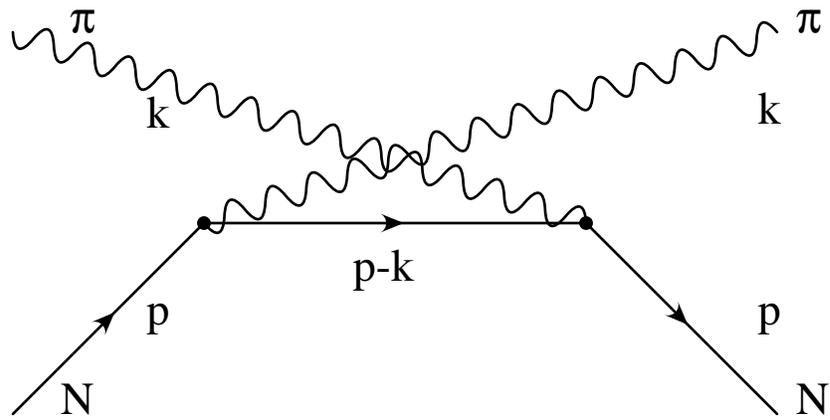
{\bf Fig. 1}. Nucleon-pole contribution to pion-nucleon
scattering.

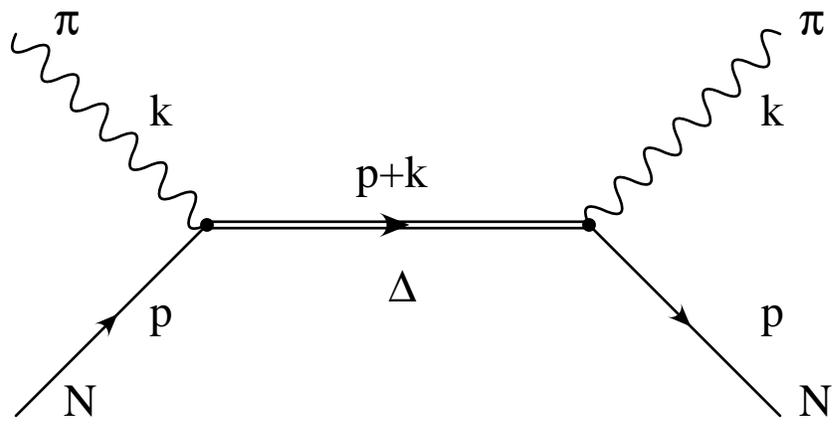
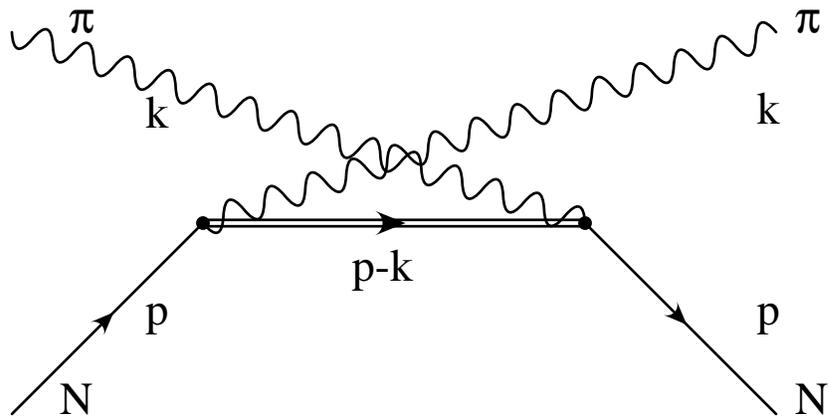
{\bf Fig. 2}. $\Delta$-pole contribution to pion-nucleon
scattering.

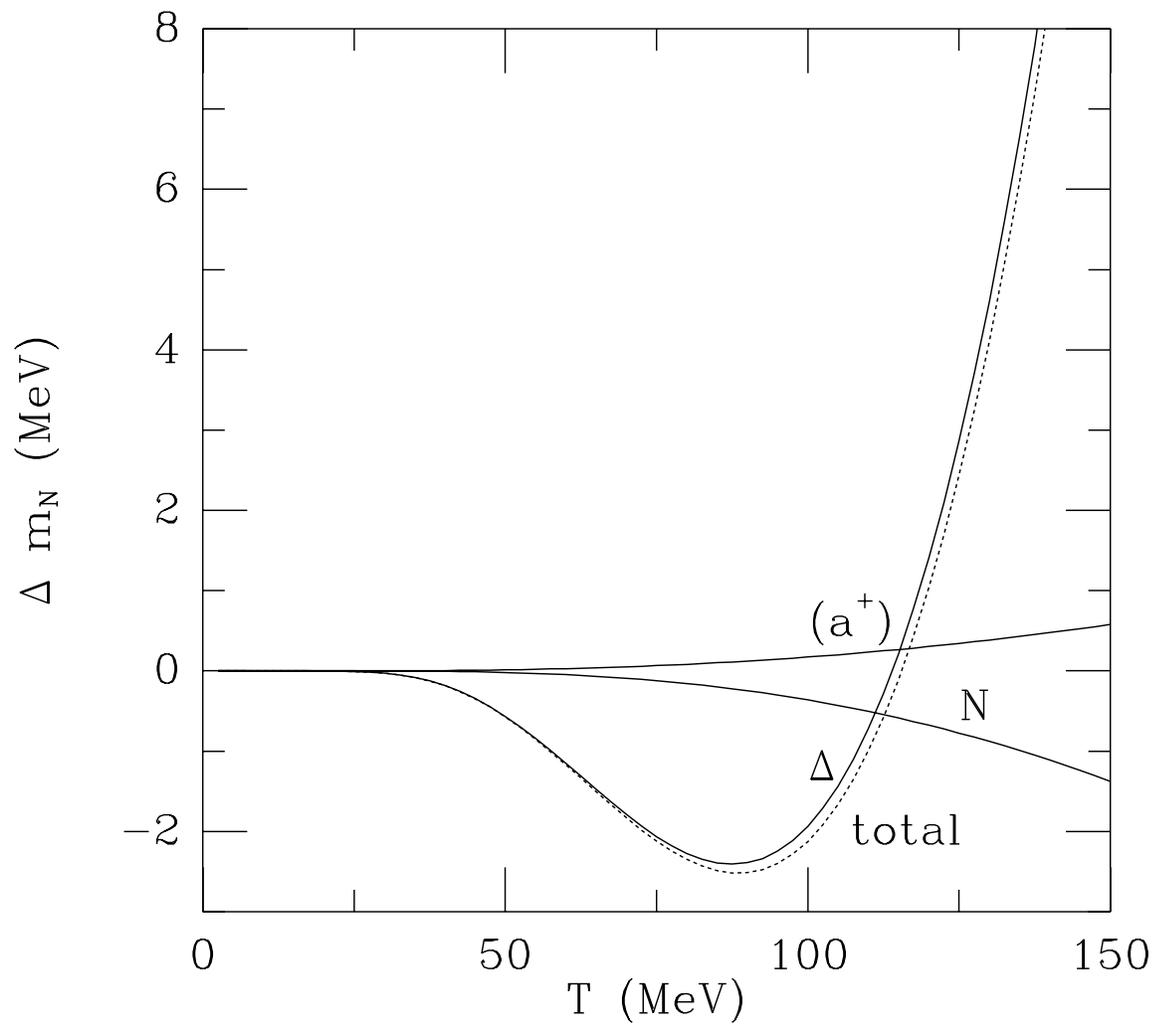
{\bf Fig. 3}. The various contributions to the nucleon mass shift
($a^{+}$, N, $\Delta$) are plotted in solid lines. The dotted line
exhibits the sum total of these contributions.

{\bf Figs. 4a and b}.  
In Figs. 4a and 4b, we respectively display in full line
the real parts ${\rm Re}\, A^{+}$ and ${\rm Re}\, B^{+}$
opposite to the experimental data points (dotted line) \cite{HOHLER}.

{\bf Figs. 5a and b}.  
In Figs. 5a and 5b, we respectively display in full line the imaginary parts
${\rm Im}\, A^{+}$ and ${\rm Im}\, B^{+}$ 
opposite to the experimental data points (dotted line) \cite{HOHLER}.

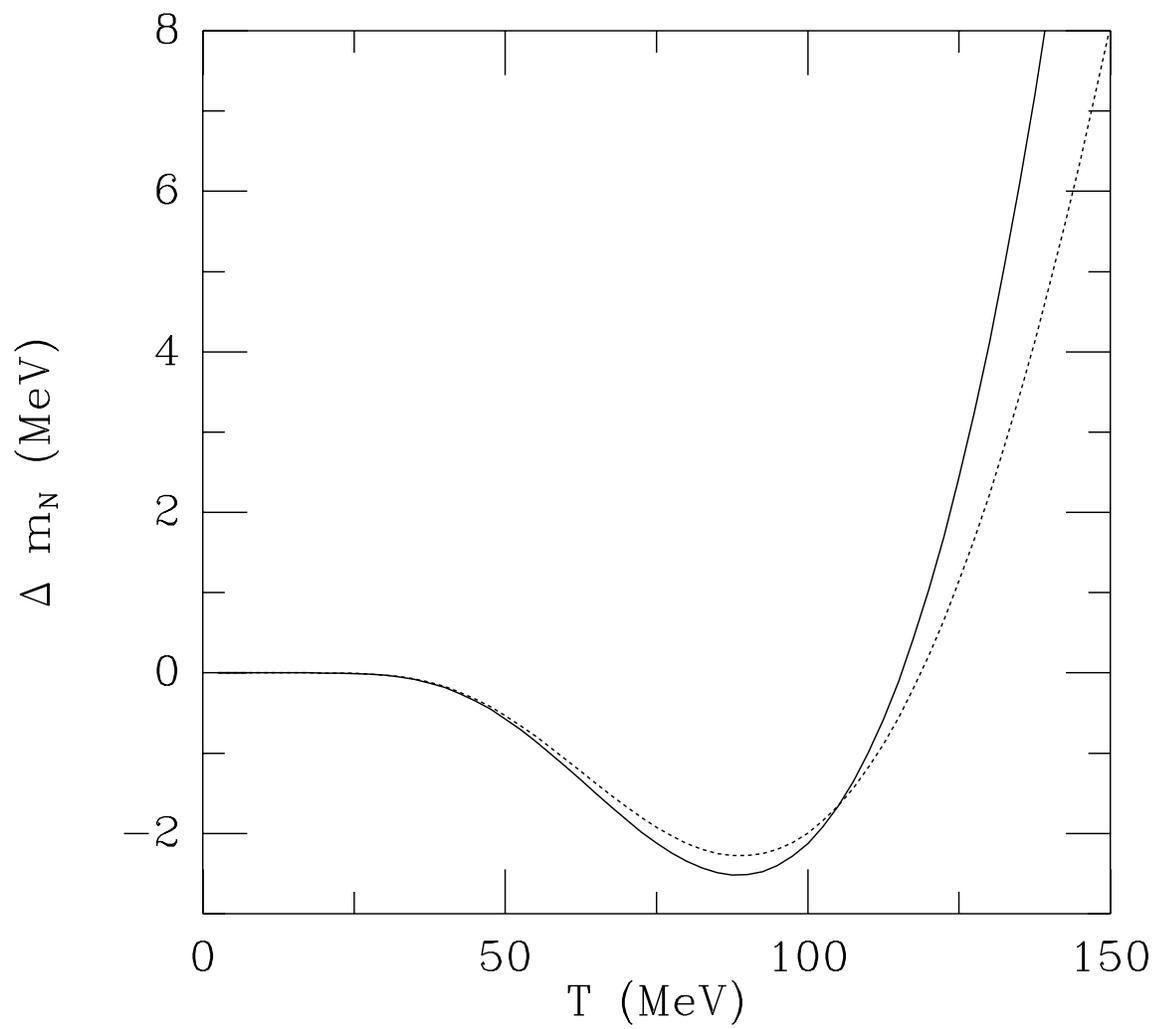
{\bf Fig. 6}. The nucleon mass shift is plotted (full line)
opposite to the one obtained from the experimental data
 points (dotted line) \cite{HOHLER}.

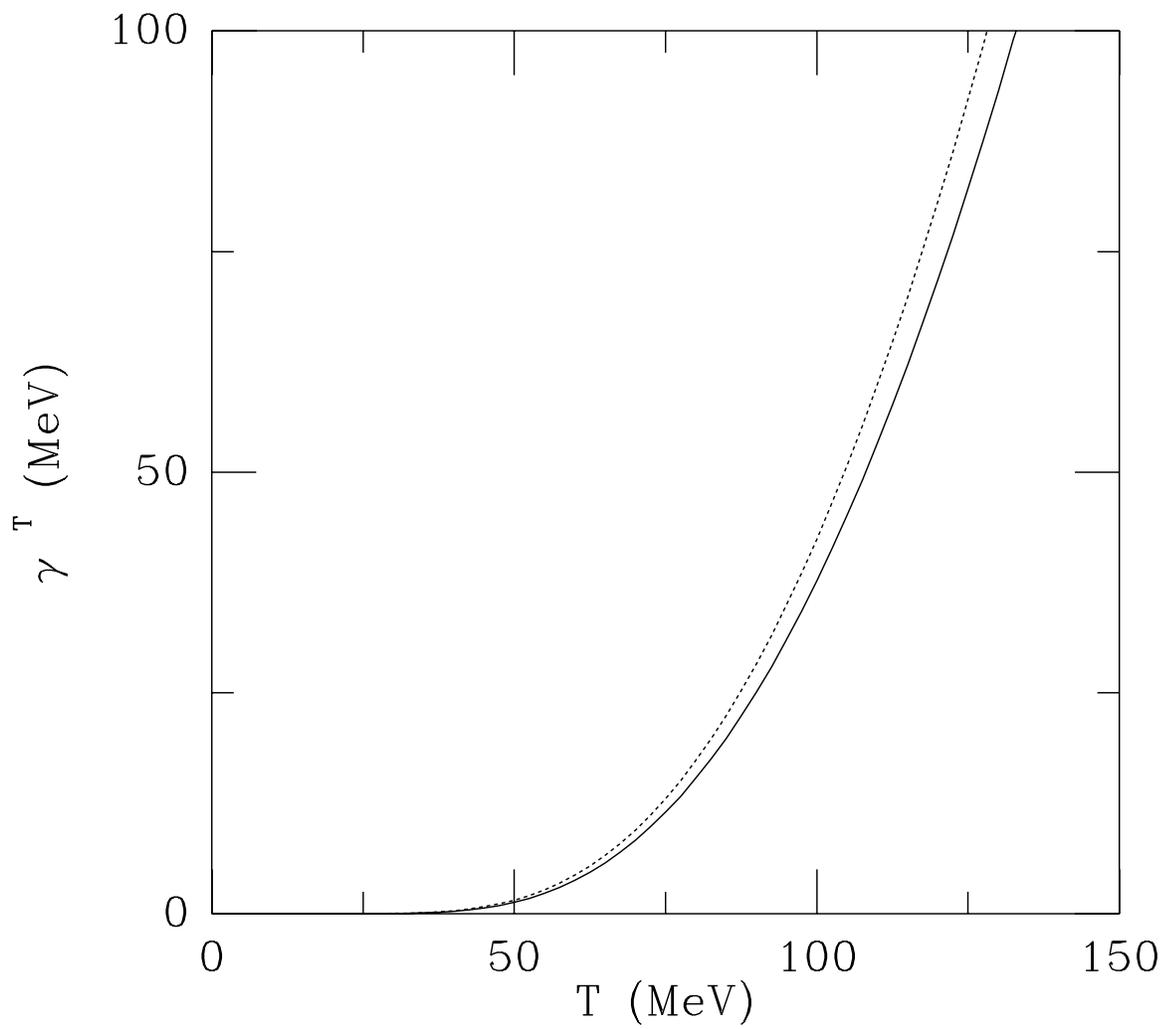
{\bf Fig. 7}. The nucleon damping rate is plotted (full line)
opposite to the one obtained from the experimental data
 points (dotted line) \cite{HOHLER}.

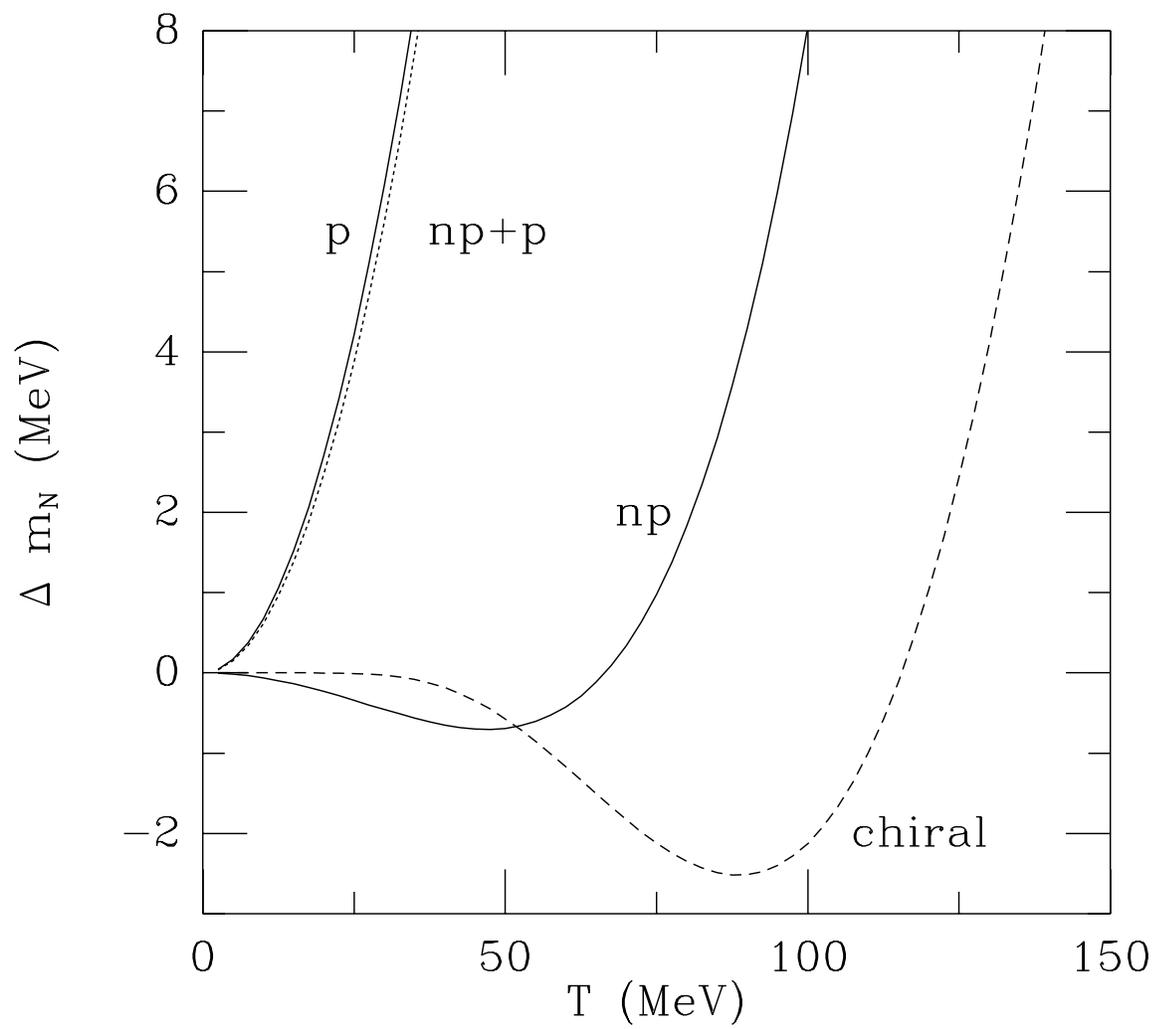
{\bf Fig. 8}. The pole (p) and non-pole (np) contributions to the 
mass shift is displayed.
 The short-dashed line represents the sum of the pole and non-pole
 contribution and the long-dashed line the mass shift from chiral
 power counting of Fig. 6.

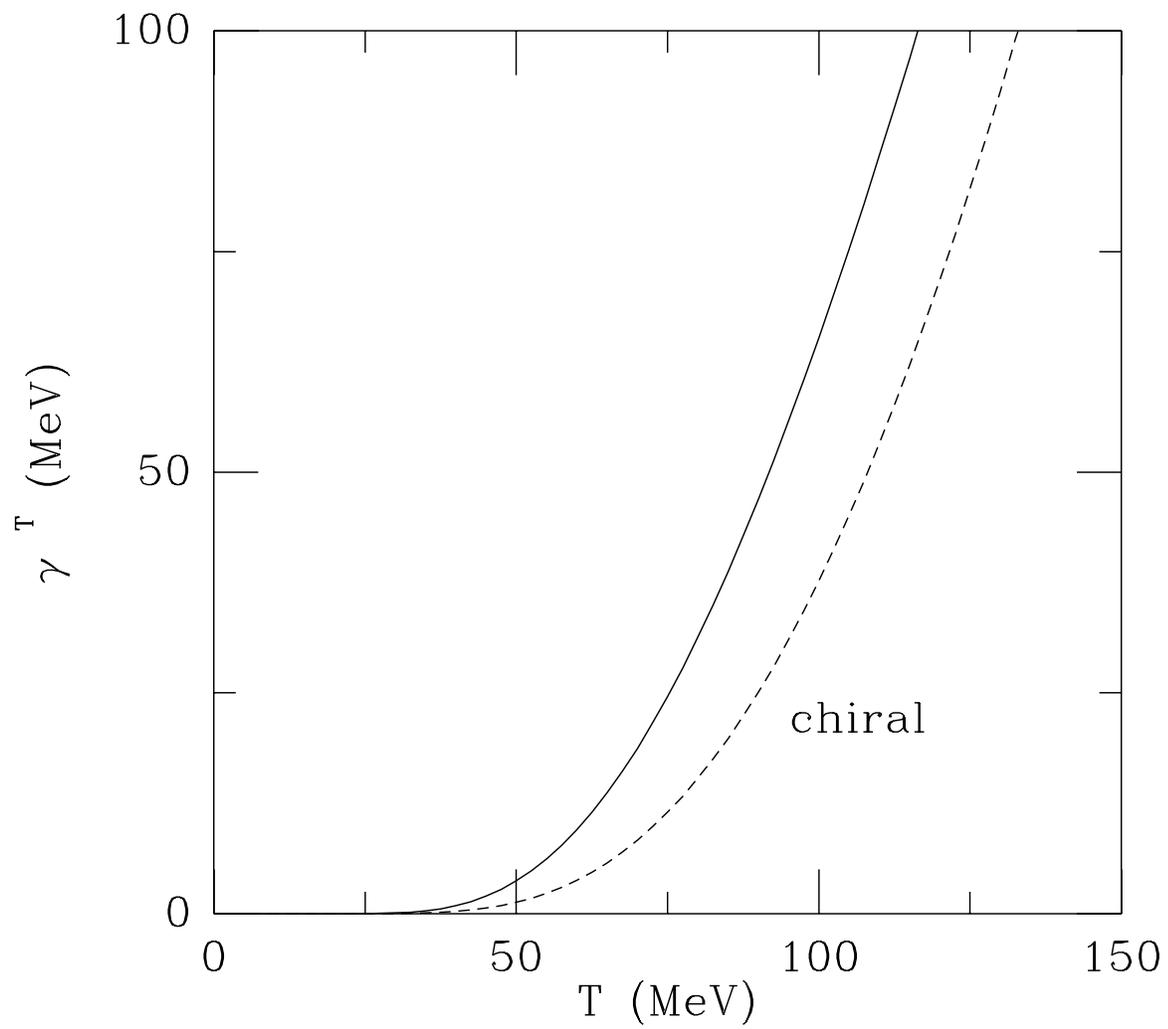
{\bf Fig. 9}. The damping rate (\ref{dpnc}) in the large $N_{c}$ 
limit is displayed in full line opposite the result obtained from 
chiral power counting (long-dashed line).

\newpage
\setlength{\baselineskip}{15pt}

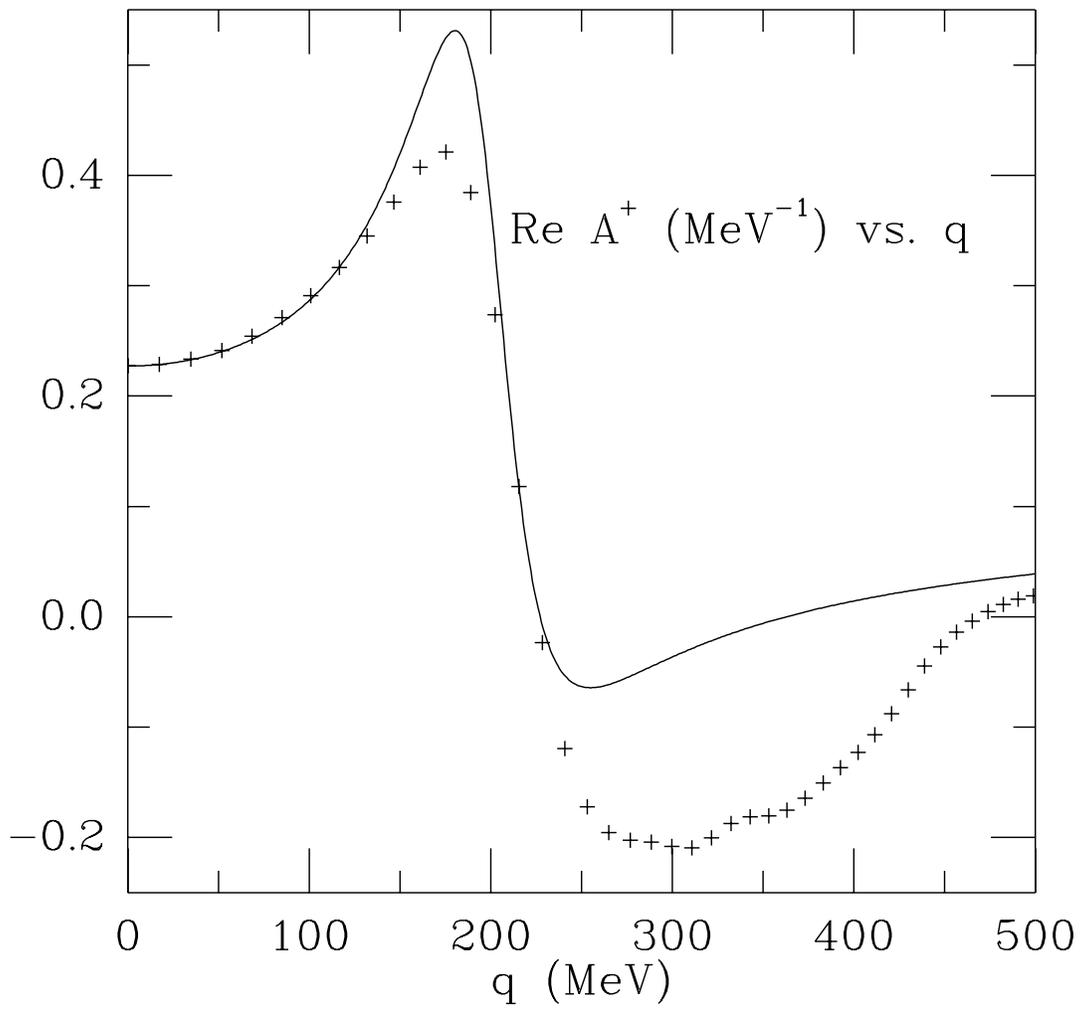
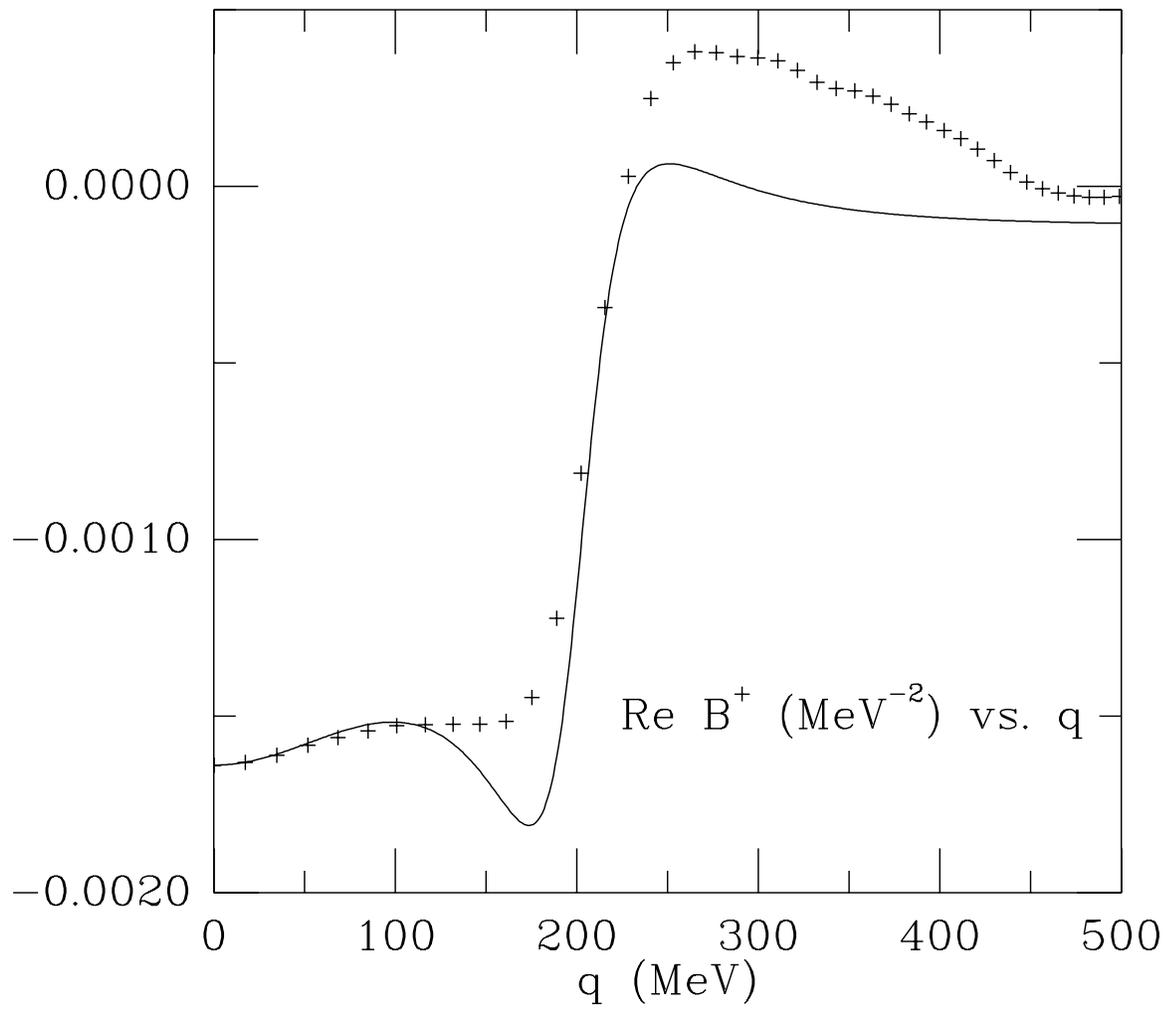
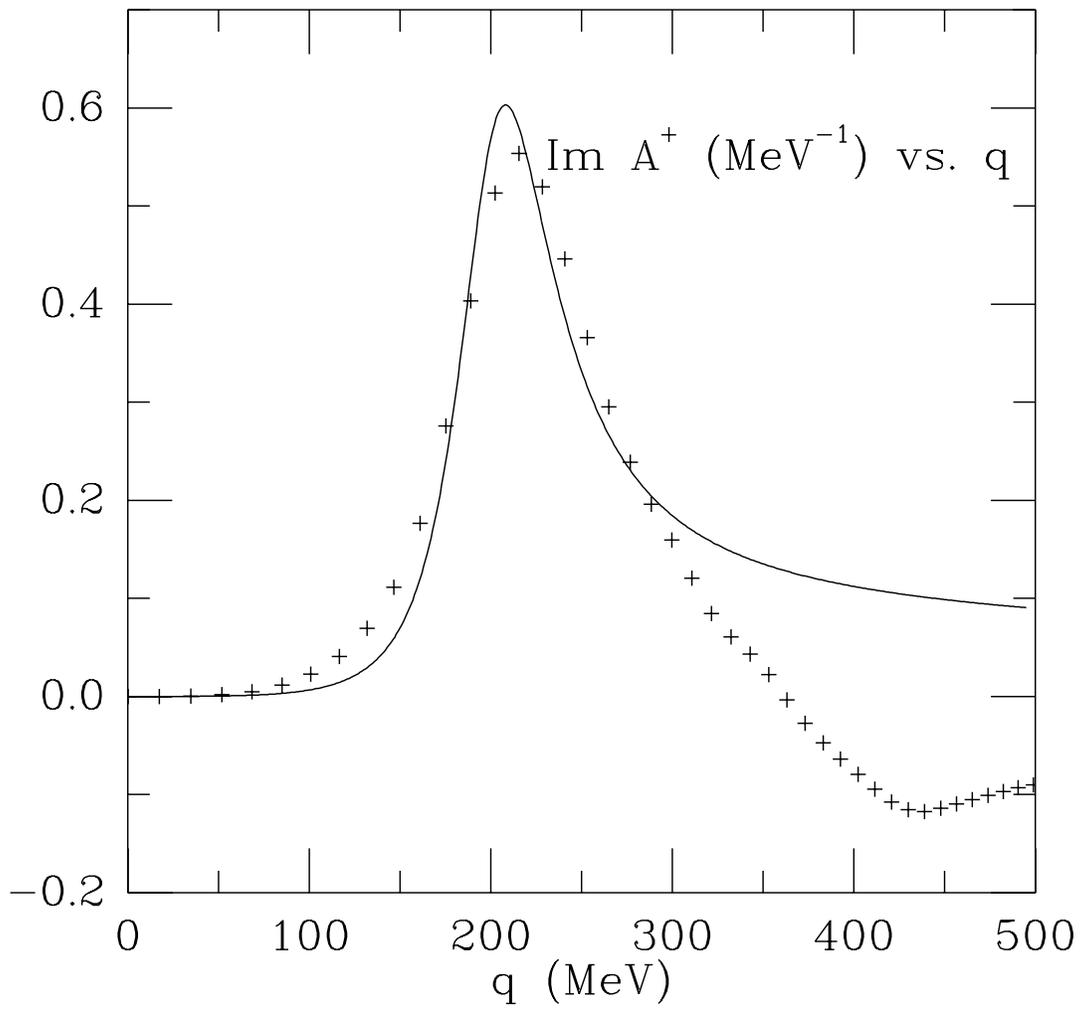
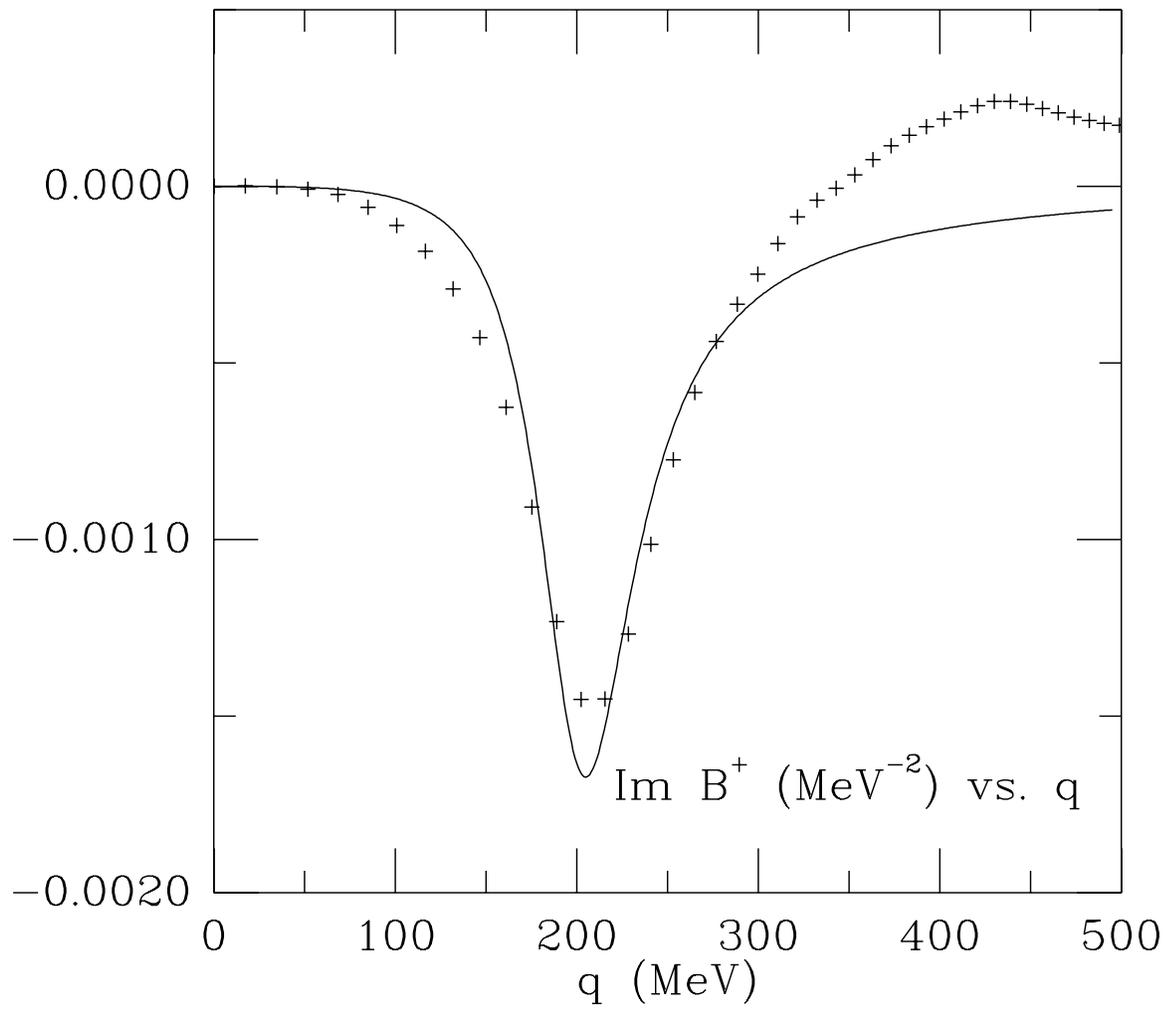
\end{document}